# High quality anti-relaxation coating material for alkali atom vapor cells

**M. V. Balabas [1,*], K. Jensen, W. Wasilewski, H. Krauter, L. S. Madsen, J. H. Müller, T. Fernholz, and E. S. Polzik**

*QUANTOP, Danish National Research Foundation Center for Quantum Optics*
*Niels Bohr Institute, University of Copenhagen, Denmark*
[1] *V. A. Fock Institute, St. Petersburg University, Russia*
[*] mbalabas@yandex.ru

**Abstract:** We present an experimental investigation of alkali atom vapor cells coated with a high quality anti-relaxation coating material based on alkenes. The prepared cells with single compound alkene based coating showed the longest spin relaxation times which have been measured up to now with room temperature vapor cells.

Suggestions are made that chemical binding of a cesium atom and an alkene molecule by attack to the C=C bond plays a crucial role in such improvement of anti-relaxation coating quality.

---

### References and links

## 1. Introduction

Alkali atom vapor cells with anti-relaxation coating are used in many experiments in quantum optics and magnetometry [1-7]. Long relaxation times of atomic polarization are necessary for these experiments, and new coating materials improving the spin relaxation time due to wall collisions "in the dark" (i.e. in the absence of optical pumping or probe light) will in many cases also directly improve the performance of these experiments. For instance, in an atomic quantum memory based on alkali vapor cells the storage time equals the transverse relaxation time $T_2$ in the dark. In atomic magnetometry the sensitivity to magnetic fields (the minimal detectable magnetic field in a given measurement time) scales as $\sqrt{1/T_2}$ .

Experience with the preparation of traditional paraffin coated cells shows that a special period of "ripening" is needed just after the preparation of the coating [11]. The ripening period is characterized by a permanent increase of atom vapor pressure up to values usually 10-15% below the saturated vapor pressure. We know that during the ripening process the coating material absorbs alkali atoms intensively. Some characteristics of absorption were measured [8, 9], but a detailed understanding of the mechanism of absorption is not established yet.

An interesting observation on the ripening process was made by performing Raman spectroscopy on paraffin coating material inside a potassium cell [10]. The spectroscopy was performed twice: just after the cell preparation and one year later. The spectral signature of unsaturated C=C bonds at $1644\,cm^{-1}$ was found in the first experiment, while no such line was detected one year later. This suggests that potassium atoms modify the molecules of the coating material during the ripening period with saturation of C=C bonds. Organometallic molecules of the type $K-CH_2(C_nH_{2n})CH_3$ are the likely reaction products formed during the ripening. As a final result we have a solid solution of organometallic molecules in a saturated hydrocarbon solvent as a chemical model of ripened anti-relaxation coating material. Such organometallic molecules should not give rise to significant extra relaxation of alkali atoms vapor ground state polarization, because these molecules have saturated bonds. An interesting question to address is how the concentration of the solid solution affects the quality of the coating.

## 2. Experiment and results

To investigate the situation we prepared coated cells with 1-octadecene ($CH_2=CH(CH_2)_{15}CH_3$) and 1-nonadecene ($CH_2=CH(CH_2)_{16}CH_3$) from Sigma-Aldrich according to the procedure discussed in [10, 11].

We measured the longitudinal relaxation time $T_1$ and the dark line width $G_0$ of the magnetic resonance for cesium cells. $T_1$ in the dark was measured by Franzen's technique [12], described briefly below. The width (HWHM) $G_0=1/(2\pi T_2)$ of the RF-optical double resonance line was measured as a function of light intensity in an actively stabilized magnetic environment at a magnetic field $B=3500nT$ and at a cell temperature of $T_{cell}=22\,^0C$ . The transverse spin decay rate in the dark is determined by extrapolation to zero light intensity.

The procedure for the $T_1$ measurement is described in detail in [17]. In brief, the relaxation time in Franzen's technique is inferred from the temporal change $S(t)$ in the absorption of circularly polarized light as a result of a change $\Delta n_i$ in the distribution of the populations of the Zeeman sublevels of the ground state due to relaxation transitions between them: $S(t) = \sum \Delta n_i(t) W_i$, where $W_i$ are the relative absorption probabilities for $\sigma^{\pm}$ polarized light. To detect the absorption signal the Cs vapor cell under study is placed in a longitudinal magnetic field and optically pumped by circularly polarized light on the $D_1$ line. Light from a low noise RF-discharge lamp, spectrally filtered by interference filters, is used for optical pumping and the transmitted pump light is measured as a function of time. Interrupting the pump light path for variable time with an electromechanical shutter (switching time below 8ms), the difference in transmission immediately before and after the dark period is used to determine the decay of the atomic polarization in the dark period.

The results of the $T_1$ measurements are presented in Fig. 1 for the case of spherical cesium vapor cells. Cell #1 has 35mm diameter and is coated with the material 1-nonadecene, cell #2 used for comparison has 40mm diameter and is coated with a standard paraffin material [11]. Fig. 1 shows the experimental Franzen signal together with a fit to a function $S(t)$ with two exponential decays $S(t) = S_1^{fast}\left[1-\exp\left(-t/T_1^{fast}\right)\right] + S_1^{slow}\left[1-\exp\left(-t/T_1^{slow}\right)\right]$. The fitted parameters are presented for standard coating material and 1-nonadecene in Tab. 1.

**Table 1. Fit parameters obtained from Franzen signal**

| | $S_1^{slow}$ [a.u.] | $S_1^{fast}$ [a.u.] | $T_1^{slow}$ [s] | $T_1^{fast}$ [s] |
|---|---|---|---|---|
| 1-nonadecene | 75.2(4.4) | 31.8(5.3) | 3.78(0.35) | 0.79(0.08) |
| Standard material | 72.0(6.1) | 26.6(6.2) | 0.25(0.01) | 0.074(0.012) |

The RF-optical double resonance line was detected by synchronous detection on an unresolved group of resonances within F=2 sublevel of the ground state of Cs atoms. Details of the setup and procedure are described in [18].

Fig. 2 shows the resonance line width as a function of photocurrent (which is proportional to light intensity) together with a linear fit for the same two cells. The dark resonance line width is given by the resonance line width extrapolated to zero light intensity and is found from the fit to be $G_0 = 1.2(0.1)Hz$ for standard coating cell and $G_0 = 0.67(0.09)Hz$ for 1-nonadecene coating cell.

Comparison of the measured parameters for the cell with the 1-nonadecene coating and a cell with almost the same characteristic size and our standard paraffin coating shows tenfold elongation of $T_1$ and 2-fold elongation of $T_2$ in the dark in favor of the 1-nonadecene coating. It should be noted that $T_2$ time measurements are sensitive to both dephasing due to wall collisions as well as to residual inhomogeneity of the applied magnetic field.

To test the observed superiority of alkene based anti-relaxation coatings a set of coated cells were prepared with several standard mixtures of alkenes from Chevron-Phillips and their laboratory-made fractions. All cells showed good characteristics. Some fraction of alkenes showed very good temperature behavior, and it was possible to get signals at $T_{cell} \geq 100\,^0C$ with characteristics only slightly worse than those at room temperature. This opens up for a wide range of coated cells for high temperature and thus high vapor density applications. Detailed results of these experiments will be reported elsewhere. Also, a few potassium and rubidium vapor cells were prepared with the new coating material. Those cells showed also very good characteristics.

To get a better understanding of the reasons for the improved performance of alkene based coatings more detailed modeling of the behavior of organometallic compounds in solid solution is desirable. Unfortunately, comprehensive information on alkali metal atoms organometallic compounds such as $K, Rb, Cs - CH_2(C_nH_{2n})CH_3$ except methyl-, ethyl-K, Rb, Cs [13] is scarce in the literature. A popular science paper by I. P. Beletskaya [14] points out that organometallic substances of alkali atoms reveal nontrivial phase equilibrium between the substances and their ionic derivatives. Each of the derivatives has its own physical and chemical properties, for example optical absorption spectrum. Equilibrium coordinates depend on various factors, almost prominently on the solvent nature and temperature. Therefore, by varying the temperature of the solution we can shift the equilibrium reversibly. Such behavior of organometallic compounds gives the opportunity to construct optimal anti-relaxation coatings for a given application.

### 3. Conclusion

We prepared vapor cells with high quality anti-relaxation coating based on the alkenes. For a well defined single compound coating material a more detailed description and understanding of alkali atom-coating interaction should be feasible. A good model for a solid solution of organometallic molecules in an alkene solvent could also be helpful to explain the results of light induced atom desorption experiments [15] and electric-field-induced density changes in coated alkali vapor cells [16].

### Acknowledgements

The authors thank the members of the Quantop research group for their enthusiastic help and discussion, and the company Chevron-Phillips for support and their material supply. M. Balabas thanks Professor A. V. Baranov for Raman spectroscopy experiments and very helpful discussions. The work was funded by the EU projects QAP, COMPASS and HIDEAS.

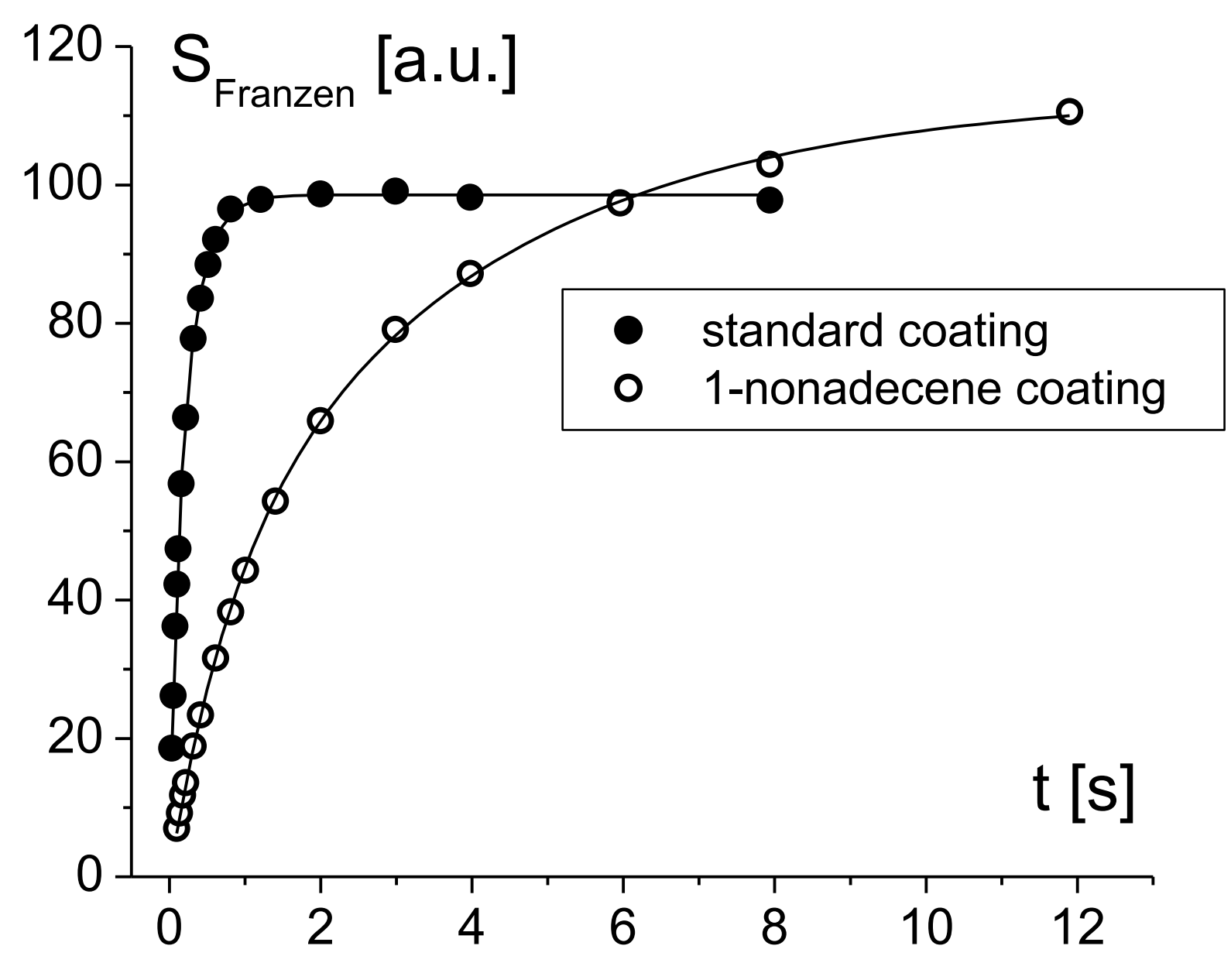

Fig.1. "Relaxation in the dark" signal with a double exponential fit

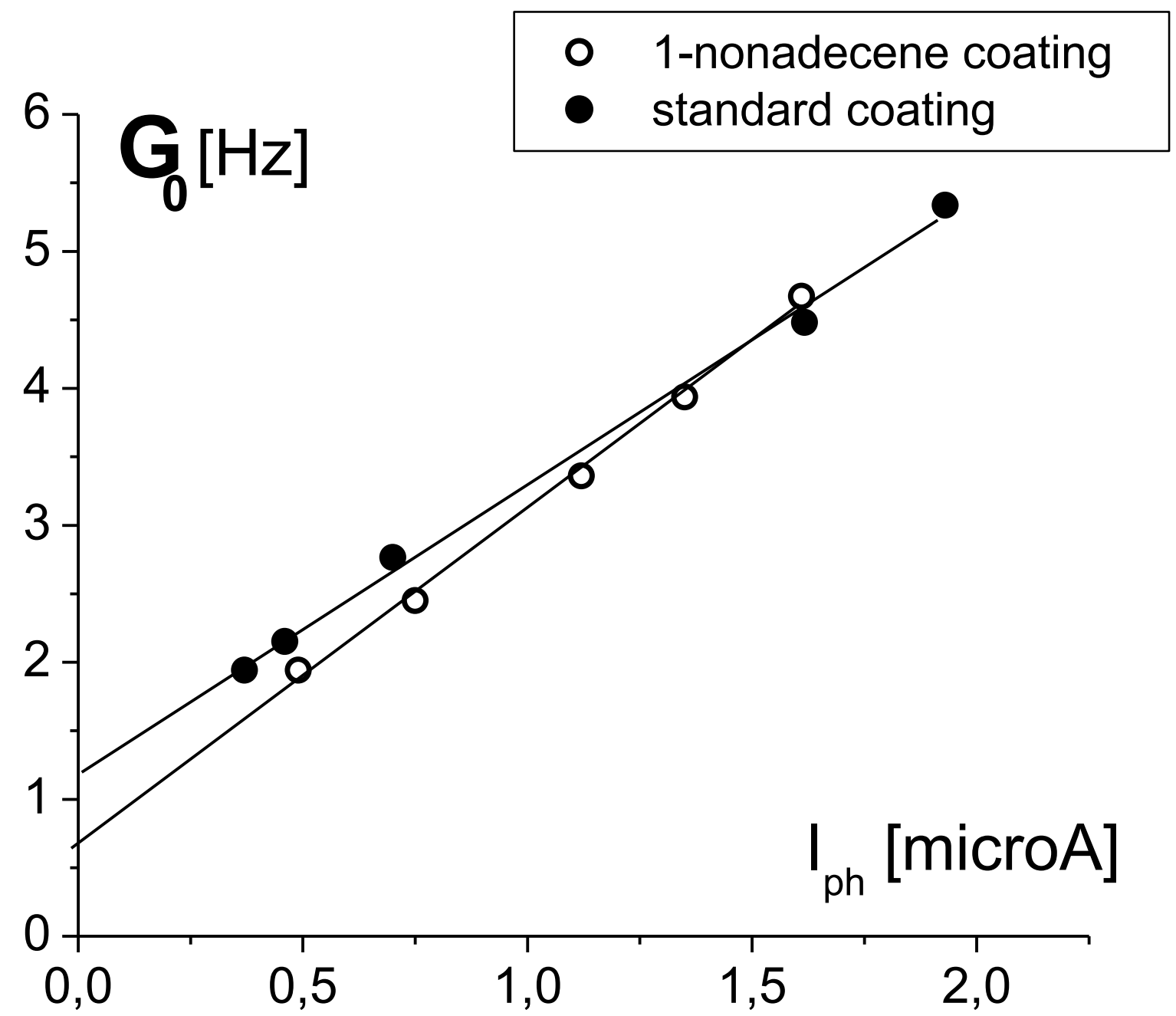

Fig.2. Double resonance line width as a function of light intensity